\documentclass[preprint]{aastex}
\usepackage{natbib}
\usepackage{booktabs}
\usepackage{threeparttable}

\begin{document}

\title{Two Kinds of Dynamic Behavior in a Quiescent Prominence Observed by the NVST}

\author{Dong~Li\altaffilmark{1,2,3}, Yuandeng~Shen\altaffilmark{2,4}, Zongjun~Ning\altaffilmark{1}, Qingmin~Zhang\altaffilmark{1,3}, and Tuanhui~Zhou\altaffilmark{1}}
\affil{$^1$Key Laboratory of Dark Matter and Space Astronomy, Purple Mountain Observatory, CAS, Nanjing 210034, People's Republic of China \\
    $^2$State Key Laboratory of Space Weather, Chinese Academy of Sciences, Beijing 100190, People's Republic of China \\
    $^3$CAS Key Laboratory of Solar Activity, National Astronomical Observatories, Beijing 100012, People's Republic of China \\
    $^4$Yunnan Observatories, Chinese Academy of Sciences, Kunming, 650216, People's Republic of China \\
    }
\altaffiltext{2}{Correspondence should be sent to: lidong@pmo.ac.cn,
ydshen@ynao.ac.cn}

\begin{abstract}
We present high-resolution observations of two kinds of dynamic
behavior in a quiescent prominence using the New Vacuum Solar
Telescope, i.e., Kelvin-Helmholtz instabilities (KHIs) and
small-scale oscillations. The KHIs were identified as rapidly
developed vortex-like structures with counter-clockwise/clockwise
rotations in the H$\alpha$ red-wing images at +0.3~{\AA}, which were
produced by the strong shear-flows motions on the surface/interface
of prominence plumes. The KHI growth rates are estimated to be
$\sim$0.0135$\pm$0.0004 and $\sim$0.0138$\pm$0.0004. Our
observational results further suggest that the shear velocities
(i.e, supersonic) of the mass flows are fast enough to produce the
strong deformation of the boundary and overcome the restraining
surface tension force. This flow-driven instability might play a
significant role in the process of plasma transfer in solar
prominences. The small-scale oscillations perpendicular to the
prominence threads are observed in the H$\alpha$ line-center images.
The oscillatory periods changed non-monotonically and showed two
changing patterns, in which one firstly decreased slowly and then it
changed to increase, while the other grew fast at the beginning and
then it changed to decrease. Both of these two thread oscillations
with changing periods were observed to be unstable for an entire
cycle, and they were local in nature. All our findings indicate that
the small-scale thread oscillations could be magnetohydrodynamic
waves in the solar corona.
\end{abstract}


\section{Introduction}
Solar prominences are one of the most common features in the solar
atmosphere. They are suspended in the tenuous hot corona and consist
of relatively dense but very cool plasma
\citep{Labrosse10,Arregui12}. Generally, the plasma in prominences
is about 100 times denser and cooler than the surrounding coronal
plasma, which raises important issues about their origin, stability,
and magnetic structures
\citep[see,][]{Mackay10,Su12,Parenti14,Cheng14,Hao15}. Solar
prominences are thought to be the most enigmatic structures
supported by coronal magnetic fields, and they are considered as the
source/driver of large-scale solar eruptions such as coronal mass
ejections (CMEs)
\citep{Okamoto07,Guo10,shen11a,shen12,Schmieder13a,Zheng17}. In
high-resolution observations, prominences are composed of numerous
thin threads \citep{Lin05,Ning09a,Yan15} that are of magnetic nature
\citep{Martin08}, and mass flows are ubiquitous along these threads.
The dynamics of these prominence fine structures might cause the
plasma instability \citep{Zirker98,Chae00,Zhang17a}. Both ground and
space observations \citep{Ning09b,Shen15,Zhang17b} showed that the
dynamic behaviors of prominences might be related to coronal
structures. Therefore, small-scale dynamic behaviors such as
Kelvin-Helmholtz instabilities \citep[KHIs,][]{Berger17,Li18} and
small-scale oscillations \citep{Okamoto07,Ning09a} are key to
understand the stability and disruption of solar prominences
\citep{Labrosse10,Arregui12,Parenti14}.

KHIs are produced by two flow-fluids in differential shearing
velocities parallel to an interface of discontinuity
\citep{Chandrasekhar61,Masson18}. They are usually observed at
strong shear-flows boundaries which can easily overcome the
restraining surface tension force and often accompanied by a process
of plasma transfer \citep{Johnson14,Berger17}. Magnetohydrodynamic
(MHD) simulations \citep{Tian16,Ni17} supported the presence of KHIs
in various coronal structures. For example, KHIs have been found in
solar twisted flux tubes \citep{Zhelyazkov12,Murawski16}, along
extreme-ultraviolet (EUV) or X-ray jets
\citep{Zaqarashvili15,Zhao18}, in solar spicules
\citep{Ajabshirizadeh15}, flares \citep{Fang16} and prominence
\citep{Antolin15}, in coronal loops \citep{Karpen94,Howson17}, in
CMEs \citep[][]{Gomez16}, and solar winds \citep{Zaqarashvili14}. In
contrast, space observations also showed evidence of KHIs in the
solar atmosphere, which are charactered by vortex-like features due
to the supersonic velocities \citep{Berger10,Johnson14}. Using the
Atmospheric Imaging Assembly (AIA) on board the {\it Solar Dynamics
Observatory} ({\it SDO}), the spatial and temporal evolutions of
KHIs were observed on the surface of fast CMEs
\citep{Foullon11,Foullon13} at a high temperature ($\sim$11~MK),
i.e., AIA~131~{\AA}. They can be found on the interface between an
erupting/dimming region and the surrounding corona in AIA EUV
passbands \citep{Ofman11}. The vortex-like structures of KHIs were
also detected on the boundary of a filament which embedded in a CME
\citep{Mostl13} or along the boundary of a twisting solar polar
coronal hole jet \citep{shen11b,Zhelyazkov18} at lower temperature,
such as AIA~304~{\AA}. Beside the {\it SDO}/AIA observations,
\cite{Feng13} reported a KHI in a coronal streamer which detected by
the {\it Solar and Heliospheric Observatory} ({\em SOHO}). Using
{\it Hinode}/Solar Optical Telescope (SOT) observations, KHIs were
observed in an active region jet \citep{Zhelyazkov16} or along the
bubble boundaries in a quiescent prominence
\citep{Ryutova10,Berger17}. All those simulations and observations
indicate that KHI is one of the magnetohydrodynamic (MHD)
instabilities and could play an important role in the dynamic of the
solar atmosphere \citep[e.g.,][]{Foullon11,Ofman11,Innes12}.

Prominence oscillations are usually classified as large-amplitude
($\geq$20~km~s$^{-1}$) and small-amplitude oscillations
\citep[$\leq$2$-$3~km~s$^{-1}$,][]{Oliver02}. Large-amplitude
oscillations are usually induced by external disturbances such as
Moreton and EUV waves, coronal jets, flares, and mini-filament
eruptions
\citep[e.g.,][]{Eto02,Chen08,Asai12,Liu13,Luna14,Xue14,Zhou17,Zhang18},
and they can affect the entire structure of prominences.
Small-amplitude oscillations in prominences are locally in nature
\citep{Thompson91,Ballester06,Soler07}, and often exhibit as thread
perturbations \citep{Lin07,Lin09,Okamoto07,Okamoto15,Ning09a}. The
oscillatory periods are observed from minutes to hours
\citep{Pouget06,Ning09b,Kim14,Shen14a}. Previous observations of
prominence oscillations usually found in the H$\alpha$ and Ca~II~H
images
\citep[e.g.,][]{Ramsey66,Jing03,Okamoto07,Zhang12,Schmieder13}, or
the He~I~584.33~{\AA} line \citep[e.g.,][]{Regnier01,Pouget06}.
Recently, the prominence oscillations have also been detected in EUV
passbands using {\it SDO}/AIA observations, such as 171~{\AA},
193~{\AA}, 211~{\AA}, and 304~{\AA} \citep{Dai12,Li12,Bi14,Shen14b}.
These studies could help us to understand their origin and physical
properties, which is one of the important issues in prominence
seismology.

High-resolution observations from the New Vacuum Solar Telescope
\citep[NVST,][]{Liu14} provides us an unique chance to investigate
the solar fine structures, such as the small-scale instabilities and
oscillations in the prominence threads. In this paper, using the
NVST and {\it SDO}/AIA \citep{Lemen12} observations, we investigate
the Kelvin-Helmholtz instabilities (KHIs) and the small-scale thread
oscillations in an off-limb quiescent prominence, i.e., S40E83. Our
observational results provide new clues to diagnose the physical
properties of the small-scale dynamic behaviors in solar
prominences.

\section{Observations and Measurements}
The NVST is an one-meter ground-based telescope located at Fuxian
Solar Observatory, whose main aim is to observe the photospheric and
chromospheric fine structures. The telescope is operated by Yunnan
Observatories of the Chinese Academy of Sciences. On 2017 September
18, a quiescent prominence located at the southeast limb of the
solar disk (S40E83) was observed by the NVST from 03:01:00 to
03:53:00 UT with the H$\alpha$ line-center and its off-bands
($\pm$0.3~{\AA}). However, the raw images taken by the NVST were
randomly and rapidly degraded due to the turbulence of the Earth's
atmosphere. Therefore, they must be firstly reconstructed by using
high-resolution imaging algorithms. In this paper, we used the NVST
level1 data which were processed by frame selection \citep[or lucky
imaging,][]{Tubbs04}. Briefly, one high-resolution image was
reconstructed from about 100 raw short-exposure images
\citep[see.,][]{Liu14,Xu14,Xiang16}. The reconstructed NVST
H$\alpha$ images have a time cadence of $\sim$25~s and a spatial
scale of $\sim$0.136\arcsec/pixel, respectively. In addition, the
EUV and H$\alpha$ images observed by the {\it SDO}/AIA
($\sim$0.6\arcsec/pixel) and GONG ($\sim$1.0\arcsec/pixel) are also
used in this paper.

\begin{figure}
\epsscale{1.0} \plotone{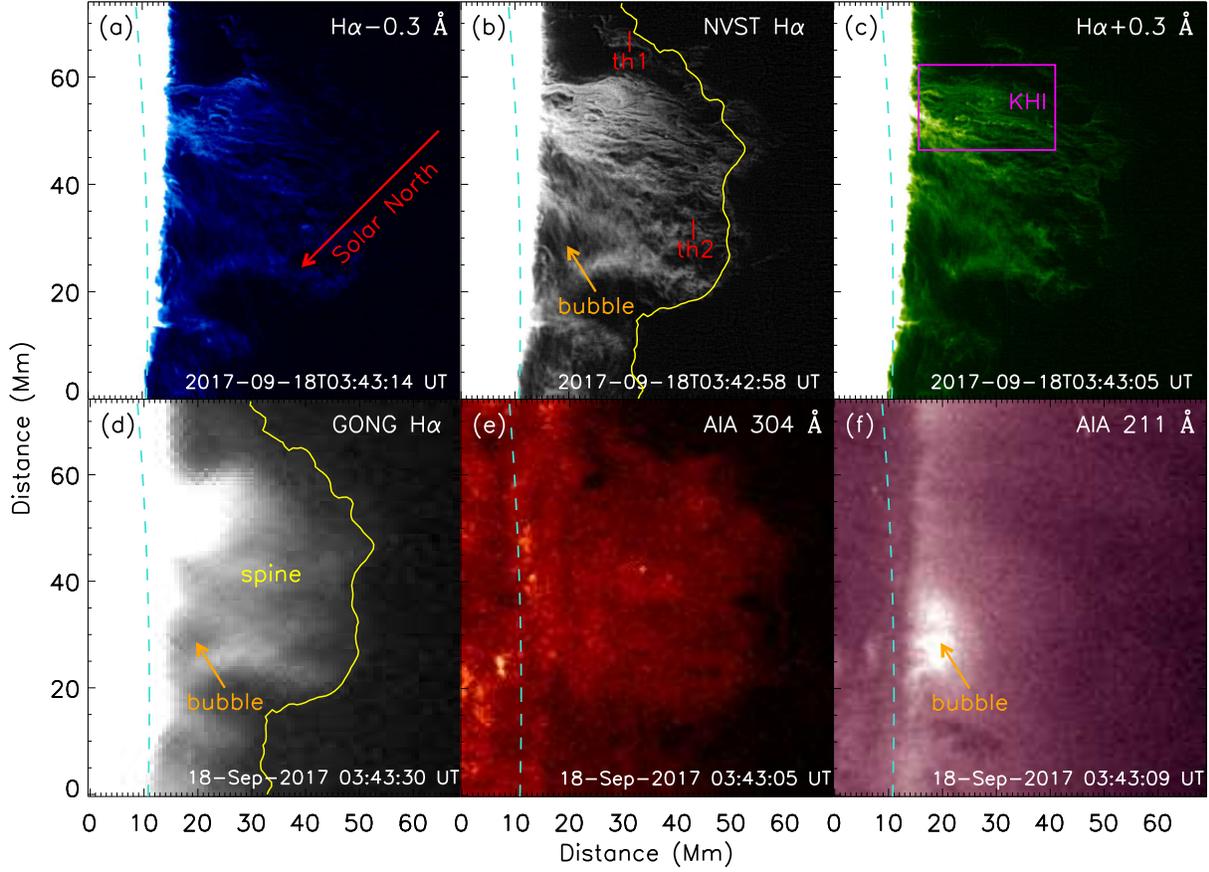} \caption{Multi-wavelength
observations show the quiescent prominence. Upper: NVST H$\alpha$
images at the line-center (b) and two extended wings at
$\pm$0.3~{\AA}  (a, c). The purple rectangle outlines the FOV in
Figures~\ref{khi1} and \ref{khi2}, two short red lines mark the
prominence threads in Figures~\ref{th1} and \ref{th2}. Bottom:
Imaging observations of GONG H$\alpha$ (d), AIA 304~{\AA} (e) and
211~{\AA} (f). The turquoise line represents the solar limb of
photosphere, the orange arrow indicates the prominence bubble, and
the yellow contour outlines the spine profile.} \label{image}
\end{figure}

Figure~\ref{image} shows the location and morphology of the
prominence in multi-wavelength images, including NVST~H$\alpha$ (b)
and its off-bands (a, c), GONG~H$\alpha$ (d), AIA~304~{\AA} (e) and
211~{\AA} (f). The prominence was `quiescent' and suspended above
the solar limb, i.e., S40E83. The main body of this quiescent
prominence shows as a typical `hedgerow' structure at the
low-temperature channels such as NVST/GONG H$\alpha$ and
AIA~304~{\AA}, which is also regarded as the prominence spine. The
spine profile is drawn from the GONG H$\alpha$ image (d), and we
also outline it in the NVST H$\alpha$ image at line-center (b), as
shown by the yellow contour. Similar to the quiescent prominence
reported by \cite{Ning09a}, the present prominence is perpendicular
to its spine axis. During our observing time interval, an intriguing
bubble formed in the prominence, which exhibited as a dark cavity in
the H$\alpha$ images (b, d) but a bright patch in the AIA~211~{\AA}
image (f) (see the orange arrow in Figure~\ref{image}). The bubble
structure can be identified in both low-(b, d) and high-temperature
(f) observational images, which is similar to the previous findings
in quiescent prominence \citep[e.g.,][]{Ryutova10,Shen15}. Based on
the NVST high-resolution observations, much more fine structures can
be observed in the quiescent prominence, such as small-scale mass
flows and thin threads (th1, th2). These mass flows and thin threads
are perpendicular to the solar limb (dashed turquoise line), which
are different from those observed in \cite{Okamoto07}, where the
authors found that the prominence was along its spine axis. However,
these fine structures are missed by GONG~H$\alpha$ and AIA EUV
images due to their lower spatial resolutions.

\section{Results}
\subsection{Kelvin-Helmholtz Instabilities}
\begin{figure}
\epsscale{1.0} \plotone{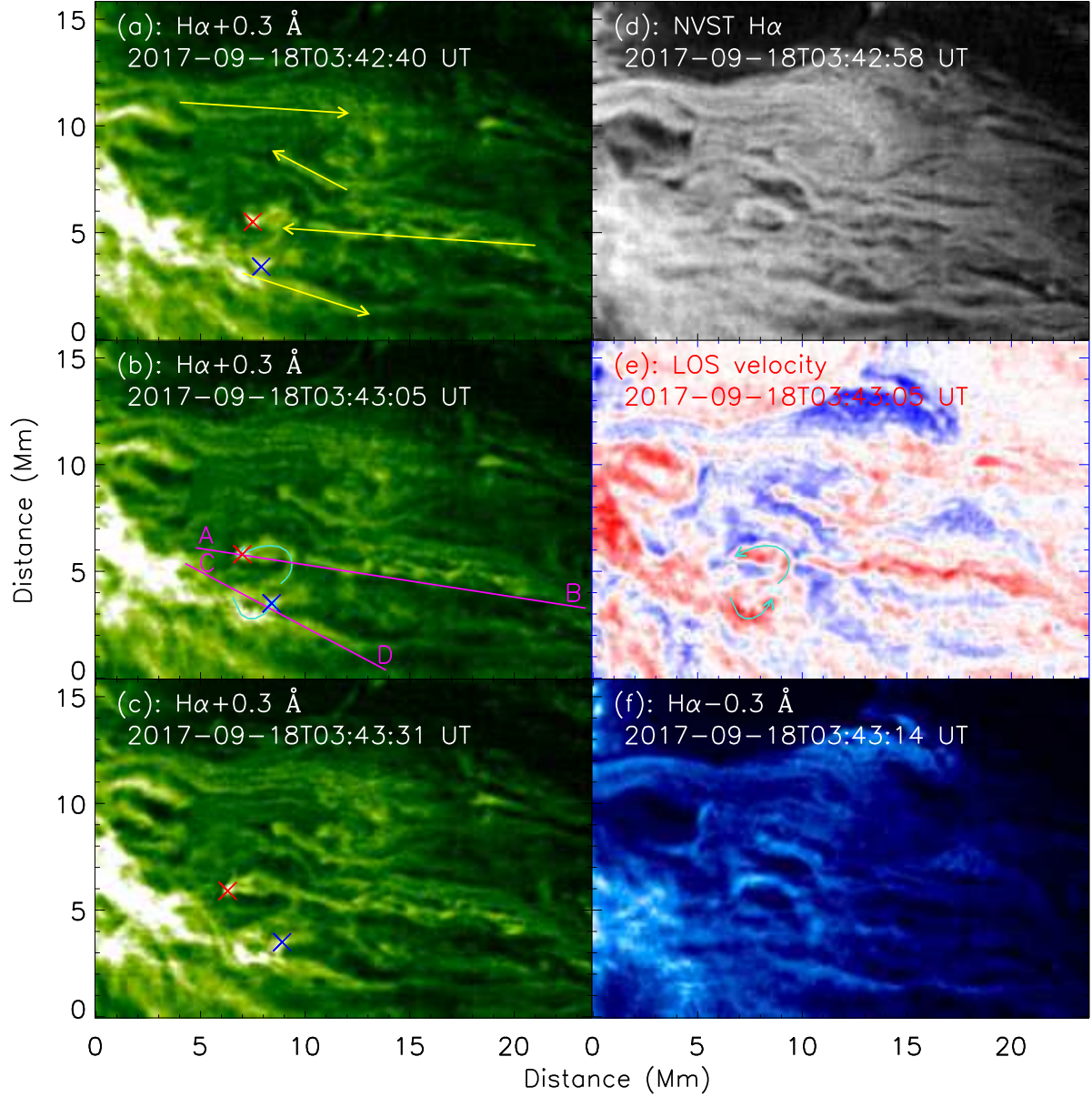} \caption{NVST H$\alpha$ images with
a small FOV outlined by the purple rectangle in Figure~\ref{image}.
Left: Time evolution of the red-wing images. The red and blue
crosses (`$\times$') mark the bright vortex-like blobs, and the
yellow arrows indicate the moving directions of plume-structures.
The purple lines A$-$B, and C$-$D are applied to draw the
time-distance images in Figure~\ref{khis}. Right: H$\alpha$ images
at the line-center (d), LOS velocity (e) and blue-wing (f). The
turquoise curved arrows mark the two vortex-like structures.}
\label{khi1}
\end{figure}

Thanks to the high-resolution observational data taken by the NVST,
we are able to investigate the fine instabilities caused by the
small-scale mass flows in the quiescent prominence.
Figure~\ref{khi1} shows the NVST images with a small field-of-view
(FOV) of around 24~Mm$\times$16~Mm, as outlined by the purple
rectangle (KHI) in Figure~\ref{image}~(c). The left panels show the
time evolution of the red-wing (+0.3~{\AA}) H$\alpha$ images at the
pronounced KHI time (b) and its nearby times (a, c). The right
panels give the H$\alpha$ images at the line-center (d) and
blue-wing (f), and also the LOS velocity image (e) between two
extended wings ($\pm$0.3~{\AA}). Those three images in the right
panels are chosen at the pronounced KHI time, i.e., around 03:43~UT.
All these NVST images show various plume-structures in the
prominence spine, no matter the H$\alpha$ line-center or the two
extended wings. The yellow arrows outline the plume-structures that
move toward to the different directions, which consist of a series
of thin and short threads. The movie of khi.mp4 further shows that
the plume-structures in the red-wing H$\alpha$ images moved at
different velocities in the plane-of-sky, which could be considered
as small-scale mass flows along the prominence plumes. When the
speeds of mass flows are fast enough to shear the boundary and
overcome the surface tension force, vortex-like structures are
formed along their surfaces/interface \citep{Johnson14}. In our
observations, two well-developed vortex-like structures were
produced on the interfaces of prominence plumes, as indicated by the
turquoise curves in panel~(b). The sizes of these two pronounced
vortex-like structures are small (say, less than 2~Mm). As indicted
by the turquoise arrows and the khi.mp4 movie, both of the two
vortex-like structures rotated counter-clockwise. In the khi.mp4
movie, the fronts of these vortex-like features were brighter than
the background prominence plasma, which displayed as bright
vortex-like blobs, as also indicated by the red and blue crosses in
the left panels. Both of these two vortex-like blobs evolved rapidly
with a time scale of minutes from the movie. Such small-scale
vortex-like blobs are most likely to be the KHIs which caused by the
strong shear-flows motions on the interface of prominence plumes.
However, theses well-developed vortex-like blobs are not found in
the line-center (d) and blue-wing (f) H$\alpha$ images. The LOS
velocity image (e) derived from two H$\alpha$ extended wings at
$\pm$0.3~{\AA} confirms this result that the vortex-like blobs only
appear pronounced at the H$\alpha$ red-wing.

\begin{figure}
\epsscale{0.9} \plotone{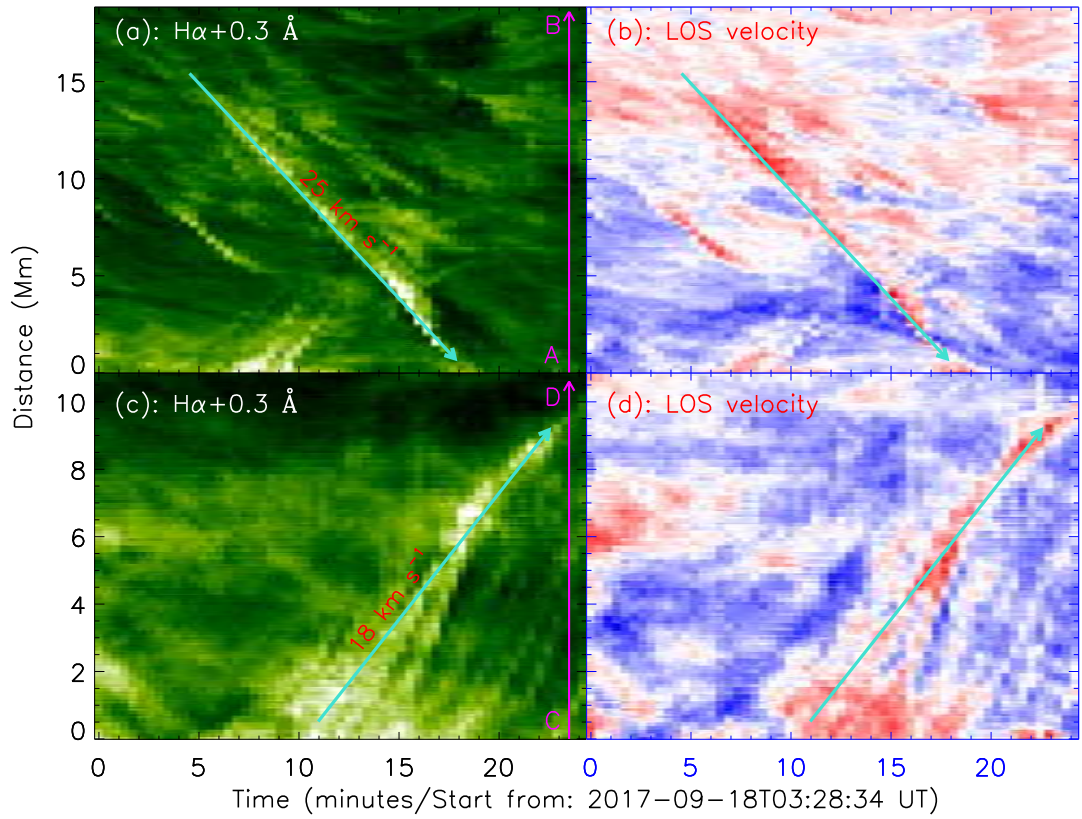} \caption{Time-distance images made
from the H$\alpha$ +0.3~{\AA} and LOS velocity images along the
purple lines in Figure~\ref{khi1}. The turquoise arrows mark the
pronounced mass flows.} \label{khis}
\end{figure}

The movie of khi.mp4 also shows that these prominence plumes moved
quickly along their interfaces. To examine the shear velocities that
were strong enough to drive the KHIs, we then plot the time-distance
(TD) images in H$\alpha$~+0.3~{\AA} along the surfaces (purple lines
in Figure~\ref{khi1}) of prominence plumes, as shown in
Figure~\ref{khis}. Panel~(a) gives the TD image along the
slit~A$\rightarrow$B, various mass flows appear on the surface of
prominence plume, and one pronounced mass flow is identified and it
is outlined by the turquoise arrow. The mass flow moves from $B$ to
$A$ at a speed of $\sim$25~km~s$^{-1}$. The appearance time of this
mass flow is consistent with the time interval of the vortex-like
structure (AB) in Figure~\ref{khi1}, and their positions are also
overlaid. All these observational results suggest that the
vortex-like structure was caused by this mass flow. The TD image in
the LOS velocity (b) also exhibit the mass flow which moves from $B$
to $A$ at a speed of $\sim$25~km~s$^{-1}$, implying that the mass
flow along the prominence plume appears obviously in the H$\alpha$
red-wing. Panels~(c) and (d) show the similar results of the other
vortex-like structure, a lot of mass flows appear on the surface of
prominence plume, and one pronounced mass flow is identified and it
is outlined by the turquoise arrow. It moves from $C$ to $D$ at a
speed of $\sim$18~km~s$^{-1}$. The appearance time and location of
this mass flow is also in agreement with the time interval of
vortex-like structure (CD) in Figure~\ref{khi1}. The measured
mass-flow speeds agree well with previous findings in prominence
plumes observed by {\it Hinode}/SOT \citep{Berger10,Shen15}. All our
findings suggest that the two vortex-like structures in the
prominence plumes could be regarded as KHIs on the interface of the
prominence plumes.

\begin{figure}
\epsscale{1.0} \plotone{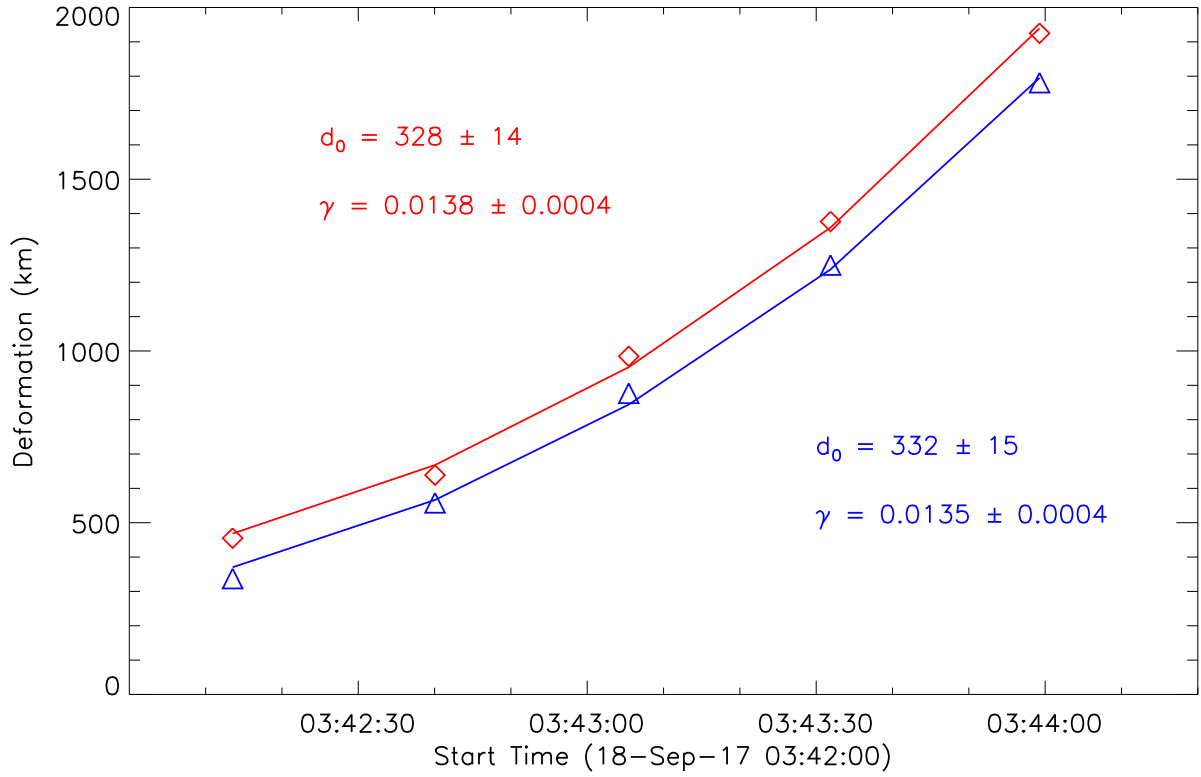} \caption{Growth rates of the KHIs.
Two bright vortex-like blobs are applied to measure the distortions
over time. The blue and red lines denote the best-fitted curves
which corresponding to the blue and red crosses in
Figure~\ref{khi1}.} \label{growth}
\end{figure}

For there to be an instability there must be the growth of a
quantity, especially in the linear phase the growth is exponential.
Therefore, the bright-blob positions which marked by the blue and
red crosses (`$\times$') in Figure~\ref{khi1} were selected to show
the quantity of vortex-like structures, such as the grow rate
($\gamma$) of KHI. Figure~\ref{growth} plots the deformation (d)
over time at two bright vortex-like blobs, i.e., blue and red
crosses. Here we assume that the KHI deformation is used as the
displacement of bright-blobs. Both of these two distortions are
growth in the exponential form. Using Equation~\ref{defo}, the
deformation is fitted with an exponential function. The blue and red
lines in Figure~\ref{growth} give the best-fitted curves for the
growth of double KHIs, and the growth rates are estimated to be
about 0.0135$\pm$0.0004 and 0.0138$\pm$0.0004, respectively. Our
results are of the same order of recent findings about the KHI
growth \citep[see,][]{Li18}.

\begin{equation}
  d = d_0 e^{\gamma t}.
 \label{defo}
\end{equation}

\begin{figure}
\epsscale{1.0} \plotone{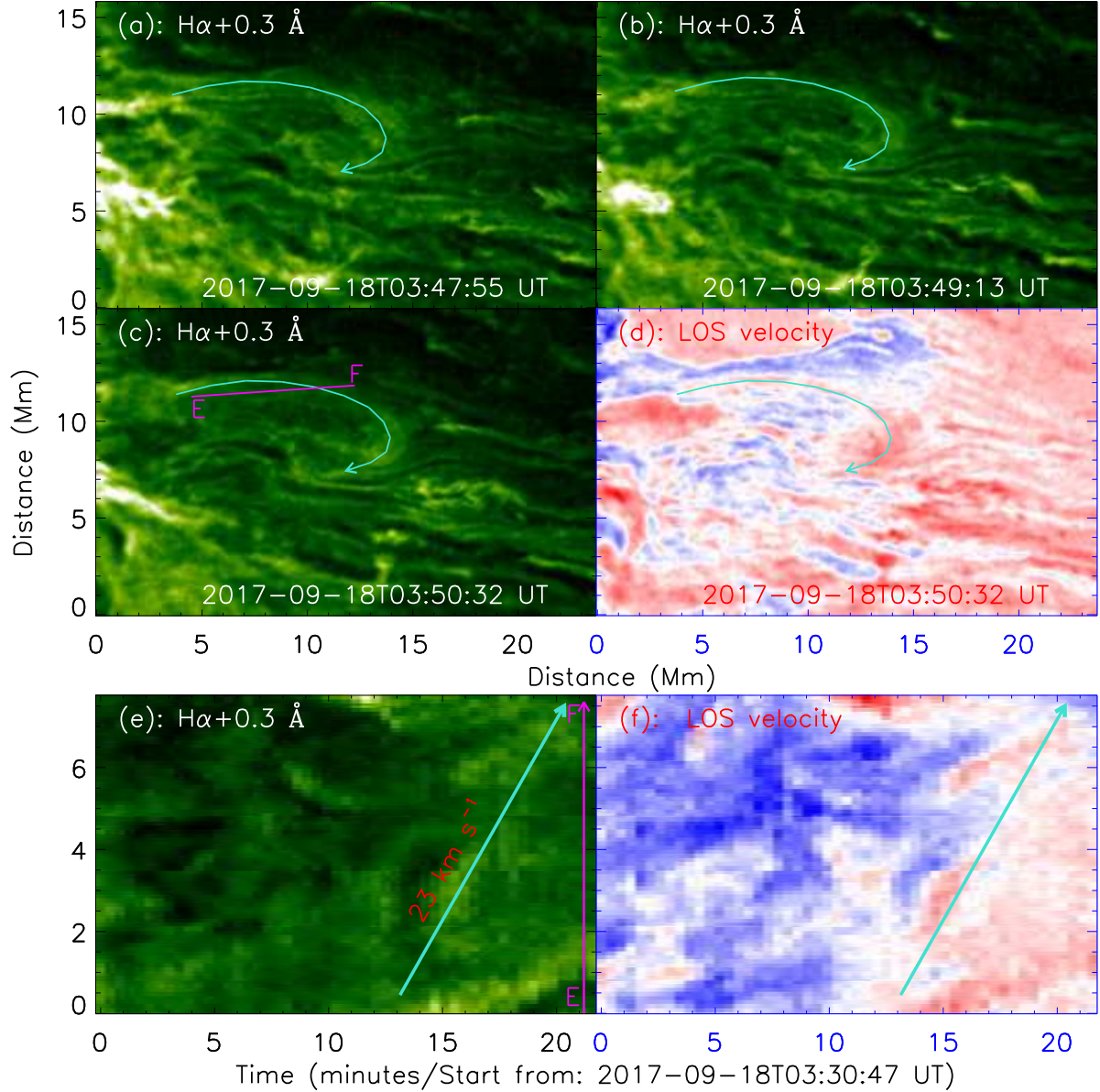} \caption{NVST images with a small
FOV (KHI). (a)-(c): Time evolution of the red-wing images. (d): LOS
velocity image. The turquoise curved arrows outline the evolution of
a vortex-like structure. (e)-(f): Time-distance images (c, d) along
E$-$F at red-wing and LOS velocity. The turquoise arrows marks the
pronounced mass flow.} \label{khi2}
\end{figure}

Figure~\ref{khi2} shows the other kind of vortex-like structure on
the surface/interface of the prominence plume, which is much larger
than those already described in the above. The FOV is the same as
that in Figure~\ref{khi1} but at a different time. Panels~(a)-(c)
show the time evolution of the H$\alpha$ red-wing images, a large
vortex-like structure (turquoise curve) appeared on the
surface/interface of prominence plumes. However, it was much weaker
than those shown in Figure~\ref{khi1}, and there was also not the
bright front in this vortex-like structure, see also the movie of
khi.mp4. The LOS velocity image (d) at the pronounced time also
confirms the vortex-like structure, and the size of this vortex-like
structure is larger ($\sim$9~Mm) than those in Figure~\ref{khi1}.
The movie of khi.mp4 and panels~(a)-(c) indicate that the
vortex-like structure rotated clockwise, as indicated by the
turquoise curved arrows. Meanwhile, the TD images at H$\alpha$
+0.3~{\AA} (e) and LOS velocity (f) along the purple line
(E$\rightarrow$F) indicates that many mass flows move on the surface
of prominence plume, and one pronounced mass flow is identified, as
outlined by the turquoise arrow. It is estimated that the speed of
the mass flow on the interface is $\sim$23~km~s$^{-1}$. The evolved
time scale is also estimated to be around 6 minutes from the movie,
i.e., from $\sim$03:46~UT till the movie end. All these
observational facts imply the appearance of a KHI on the surface of
quiescent prominence. Notice that we only plot the red-wing
+0.3~{\AA} and its LOS velocity image, because the KHIs in this
quiescent prominence were only pronounced in the red-wing H$\alpha$
images.

\subsection{Small-scale Oscillations}
\begin{figure}
\epsscale{1.0} \plotone{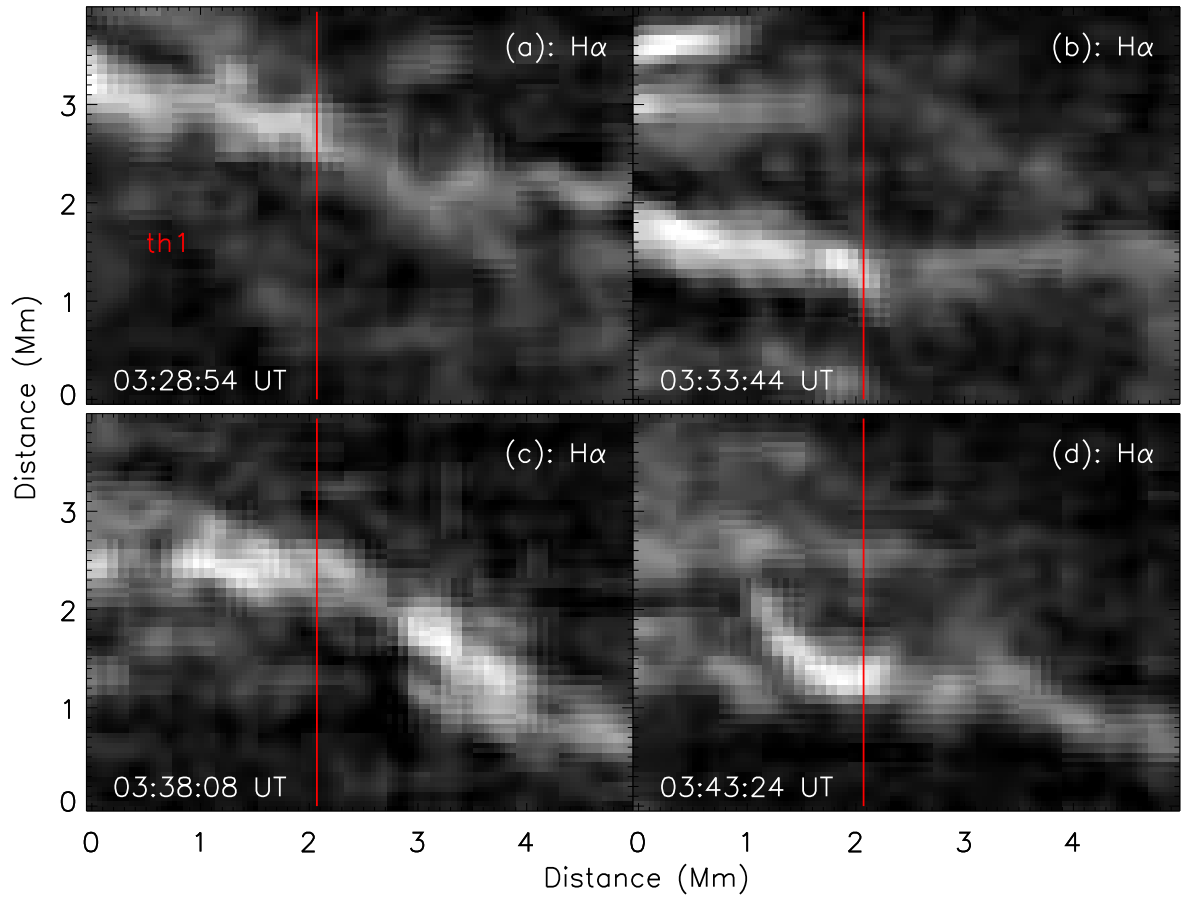} \caption{Time evolution of the
prominence thread (th1) in the NVST H$\alpha$ line-center images.
The red line marks the pronounced oscillatory position that is used
to plot the TD images in Figure~\ref{os1}.} \label{th1}
\end{figure}

Benefited from the high-resolution H$\alpha$ observations take by
the NVST, the fine thread structures with sub-arcsec width in the
quiescent prominence can well be observed, such as th1 and th2.
Figure~\ref{th1} shows the evolution of a thin and short prominence
thread (th1, case~1) in the H$\alpha$ line-center images. The thread
width is estimated to be $\sim$600~km, which is similar to previous
findings obtained by {\it Hinode}/SOT
\citep{Lin05,Okamoto07,Ning09a} or NVST \citep{Shen15}. The thin
thread exhibited a periodic movement, and a complete period is shown
in these four H$\alpha$ images in Figure~\ref{th1}. For example,
panels~(a) and (c) show that the thin thread located at
$\sim$2.5~Mm, which indicate that the positions are close to the
peak times. While it sited at $\sim$1.7~Mm in panels~(b) and (d),
suggesting the positions around the lower times.

\begin{figure}
\epsscale{0.9} \plotone{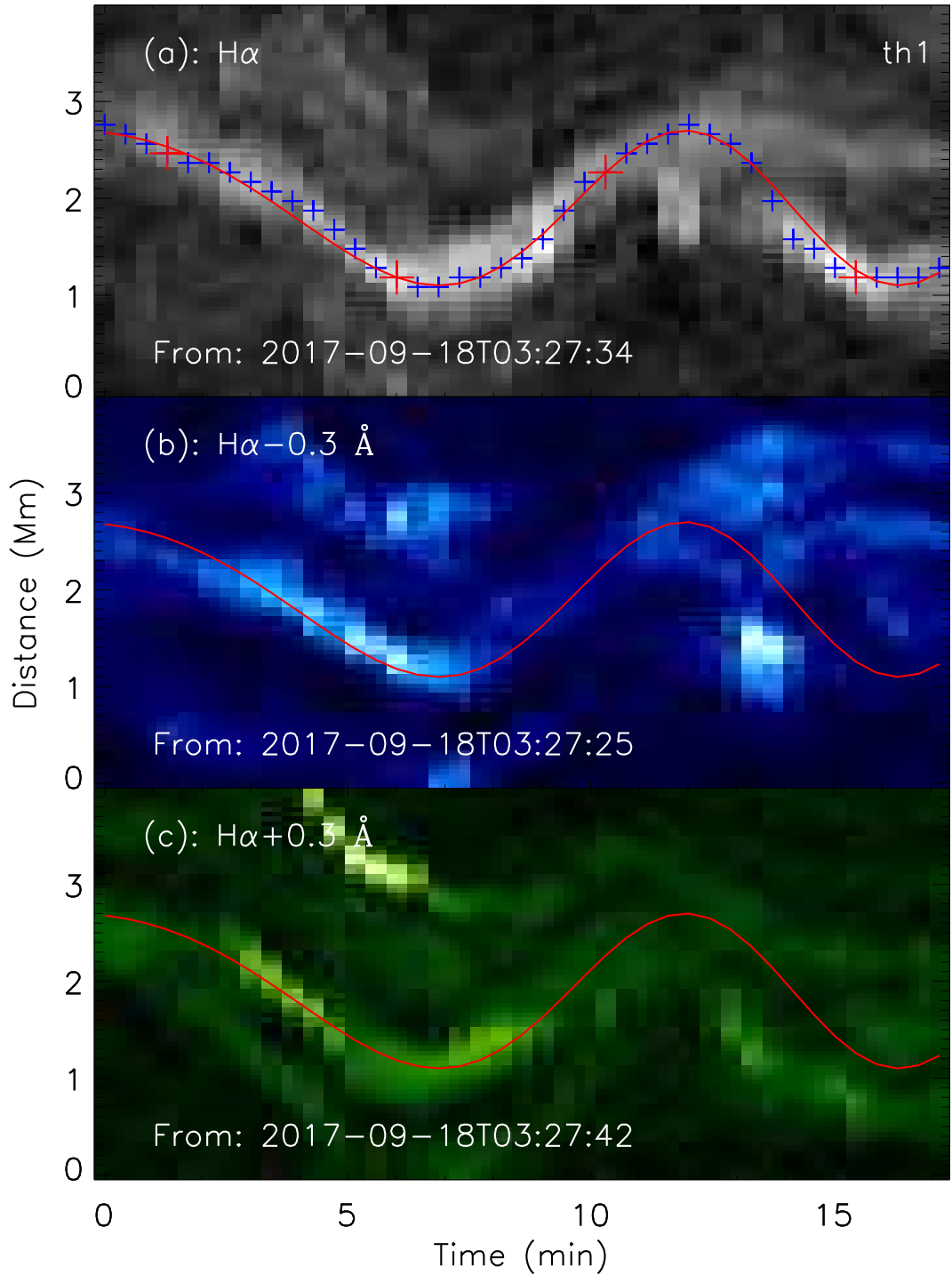} \caption{Time-distance images taken
from the red-line positions in Figure~\ref{th1} at the H$\alpha$
line-center (a) and two extended wings (b, c). The pluses (`+')
outline the brightest pixel of prominence thread, the red curve
gives the best fit of these points. The red pluses mark the times of
prominence thread given in Figure~\ref{th1}.} \label{os1}
\end{figure}

To examine the period of the thread oscillations, we plot TD images
along the red line which is perpendicular to the thin thread, as
shown in Figure~\ref{os1}. Panel~(a) shows the TD image in the
H$\alpha$ line-center, which exhibits a pronounced oscillatory
behavior. The brightest pixels in the prominence thread are selected
at each given time, as marked by the blue and red pluses (`+'). Here
the red four pluses outline the times of the images shown in
Figure~\ref{th1}. We note that they are around the peak/lower times.
Next, we fit these selected points (including blue and red pluses)
with a sinusoidal signal \citep[e.g.,][]{Zhang17a,Zhang17b,Su18}.
Equation~\ref{yfit} shows the fitting function, and the oscillatory
period (P) is changing with time, which is a second order
polynomial, indicting a non-monotonic changed period.

\begin{equation}
  A(t) = A_0+k_0 t + A_m \sin(\frac{2\pi t}{P_0+k_1 t+k_2t^2} + \phi).
 \label{yfit}
\end{equation}
Here A$_0$ is the initial position, k$_0$ indicates the thread
drifting speed, A$_m$ is the amplitude of the thread oscillations,
$\phi$ represents the initial phase shift. The oscillatory period is
a function of time, P$_0$ is the initial oscillatory period, k$_1$
and k$_2$ indicate the changing (decreasing/growth) rates of the
oscillatory period. Finally, an initial oscillatory period of
$\sim$20~minutes are derived from this thin thread, with a changing
rates of $-$0.9 and +0.022 (table~\ref{yfit}). Notice that `$-$'
indicates the oscillatory period is decrease, while `+' implying a
growth rate. Considering the initial period of $\sim$20~minutes, the
changing rates suggest that the oscillatory period first decreased
slowly and then it changed to increase. The oscillatory amplitude is
also estimated to be around 800~km. This small-scale oscillatory
phenomenon was pronounced in H$\alpha$ line-center images (panel~a),
but it was very weak and even disappeared after a half cycle in the
two H$\alpha$ wings, such as H$\alpha~\pm$0.3~{\AA}, as shown in
panels~(b) and (c).

\begin{figure}
\epsscale{1.0} \plotone{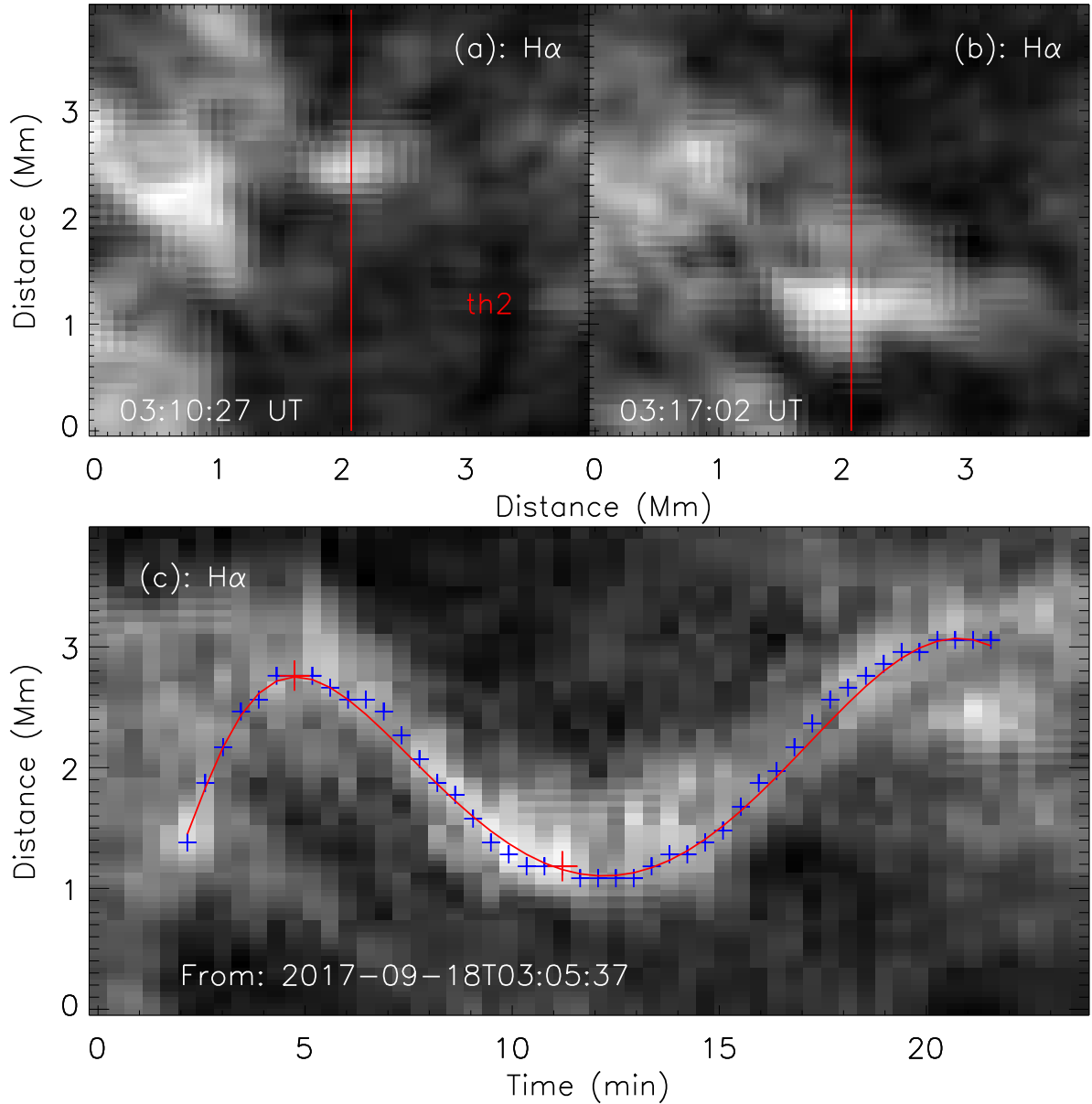} \caption{Upper: The evolution of the
other prominence thread (th2) at NVST H$\alpha$ line-center. Bottom:
Time-distance images taken from the red-line positions in panels~(a)
and (b) at the H$\alpha$ line-center. The color pluses and curve are
the same as that in Figure~\ref{os1}.} \label{th2}
\end{figure}

Figure~\ref{th2} shows another case (case~2) of thread oscillations
in the quiescent prominence. The upper panels show another thin
thread (th2) at two different times, i.e., around the times of peak
(a) and lower (b) positions, as marked by the red pluses in the
bottom panel. The blue and red pluses outline the brightest pixels
of the prominence thread too. We apply the same sinusoidal function
(Equation~\ref{yfit}) to fit these points, and obtain an initial
oscillatory period of $\sim$6~minutes, with a changing rates of +0.9
and $-$0.025 (table~\ref{yfit}), which indicates that the thread
oscillatory period grew fast at the beginning, then it changed to
decrease. The oscillatory amplitude is estimated to be about 900~km.
This thread showed a drifting motion at a slow speed of
$\sim$0.33~km~s$^{-1}$ during the oscillation, similar to previous
findings as reported in \cite{Ning09a}. The small-scale oscillations
were also only observed in the H$\alpha$ line-center images.

\section{Discussions}
The small-scale mass flows along the interface/surface of prominence
plumes always move at different speeds. The flow speeds are
estimated to be at about 25~km~s$^{-1}$, 18~km~s$^{-1}$, and
23~km~s$^{-1}$, respectively. Assuming that the temperature of the
quiescent prominence is around 7000~K \citep{Hirayama85}, the sound
speed in the quiescent prominence can be estimated as
$v_s~\sim~147\sqrt{T/MK}~\thickapprox$~12~km~s$^{-1}$
\citep{Aschwanden05}, the measured mass flow speeds are larger than
the local sound speed, which indicate the flow speeds are of
supersonic \citep{Berger10}. Thus, the shear-flows speeds that along
the interface/surface of prominence plumes are strong enough (i.e,
supersonic) to deform the boundary and overcome the restraining
surface tension force. This suggest that KHIs can occur on the
interface/surface \citep{Chandrasekhar61,Johnson14,Masson18}. Our
observational facts support this idea very well. Pervious studies
found that KHIs are important in dissipation of free energy in the
shearing flows and plasma heating \citep{Karpen94,Ofman11,Cavus13}.
Therefore, the KHIs could play an important role in the process of
energy transfer in prominence plasmas.

It is very interesting that the KHIs in the quiescent prominence
only appeared clearly in the red-wing H$\alpha$ at +0.3~{\AA}
(Figures~\ref{khi1} and ~\ref{khi2}). So far, most observations of
KHIs in the solar atmosphere are at the surface of CMEs
\citep{Foullon11,Foullon13} or jets \citep{shen11b,Zhelyazkov16}. By
using observational data taken by the {\it Hinode}/SOT H$\alpha$
line-center or Ca~II~H lines,  \cite{Berger10} and \cite{Ryutova10}
showed that prominence plumes were often highly turbulent and they
are apt to form vortex structures (i.e., KHIs) on the interface
between prominence and corona, such as along the boundaries of
prominence bubbles. In the present paper, we detect the vortex-like
structures on the interface/surface of prominence plumes in the
red-wing H$\alpha$ observations at +0.3~{\AA} shifted from
line-center, which indicate that the observed small-scale mass flows
or the vortex structures in the prominence were moving away from us.

It is also very interesting that the small-scale oscillations with
changing periods are detected in the fine threads that compose of
the quiescent prominence. The amplitudes of thread oscillations are
less than 1000~km, such as 800~km and 900~km. Thanks to the high
spatial resolution ($\sim$100~km/pixel) of NVST, such
small-amplitude oscillations can be observed. The alignment of NVST
images can be as accurate as one pixel. Therefore, the small-scale
oscillations are reliably. As the prominence threads are of magnetic
nature \citep{Martin08}, the changing periods might be caused by the
variations of the inherent physical properties of the thin threads
such as the magnetic field strength and the plasma density. The thin
threads exhibited two kinds of oscillation behaviors: the one (th1)
decreased its period with a slow rate, while the other one (th2)
increased its period with a fast rate. However, the oscillatory
amplitude did not decay in time, which is different from the
large-scale prominence oscillations that usually show strong damping
\citep[e.g.,][]{Ning09b,Zhang12,Shen14b,Zhang17a}. To the best of
our knowledge, this is the first report of the small-scale
individual oscillations in prominence threads that changed their
periods (growth/decreasing) with time \citep[see,][]{Ning09a}.

Both of the two oscillating prominence threads can last for one
entire cycle, which are similar to previous findings about the
small-scale thread oscillations
\citep{Okamoto07,Okamoto15,Lin09,Ning09a}, but different to
large-scale prominence oscillations that usually last for several or
even dozen of cycles \citep[e.g.,][]{Li13,Shen14a,Zhang17a}. The
oscillating amplitudes are less than 1000~km, which indicate that
the drivers of those thread oscillations were of small scale and
might be numerous \citep[see
also,][]{Okamoto07,Okamoto15,Ning09a,Ning09b}. These small-scale
thread oscillations are only observed pronounced in the H$\alpha$
line-center images, but missed at the two extended wings
(Figure~\ref{os1}). This suggests that there are no strong
upflows/downflows in these prominence threads.

Finally, the Sun was quiet on 2017 September 18, only one active
region (N08W39) appeared on the solar disk, but it was far away from
the solar prominence (S40E83). Moreover, we could not find any
small-scale eruptions around this quiescent prominence. Therefore,
the small-scale oscillations of prominence threads were not caused
by external disturbances such as flares or other kinds of eruptive
activities on the Sun as what has been reported in previous studies
of large-scale prominence oscillations
\citep[e.g.,][]{Jing03,Isobe06,Chen08,Shen14a,Shen14b,Zhang17b}. Our
findings support the scenario that small-scale thread oscillations,
which are perpendicular to the prominence threads, are local in
nature, and the driving of such kind of oscillations could be MHD
waves that stem from the photospheric magnetic field
\citep{Joarder97,Diaz05,Okamoto07,Lin09,Ning09a}, such as kink mode
due to transverse displacement of  thin threads in solar prominences
\citep{Edwin83,Terradas08}.

The small-scale KHIs and thread oscillations are simultaneously
observed in a same quiescent prominence, but they are independent
dynamic behaviors in the fine prominence structures. Because that
they only appeared clearly in the H$\alpha$ red-wing (KHIs) and
line-center (oscillations) H$\alpha$ images, respectively. The
correlation-ship between their temporal and spatial relationships is
also not found. Therefore, the two kinds of dynamic behavior
detected in the prominence were independent to each other, which is
different to previous findings that the KHI at the thread boundary
is triggered by transverse oscillations \citep{Okamoto15}.

\section{Summary}
Two kinds of dynamic behavior of the fine structures in a quiescent
prominence is studied in detailed based on the high-resolution
observational data taken by the NVST. The primary results of this
study are summarized as following:

\begin{enumerate}
\item The KHIs in a quiescent prominence are detected on the
interface/surface of prominence plumes. They are identified as
vortex-like structures with rapidly rotation motions in the
H$\alpha$ red-wing images, but missed by the H$\alpha$ line-center
and blue-wing observations.

\item The KHIs exhibited two kinds of dynamic behavior in the same
prominence. One rotated counter-clockwise, and it showed as
small-scale bright vortex-like structures ($<$2~Mm). The other one
rotated clockwise, and it showed as relatively larger but weak
vortex-like structure ($\sim$9~Mm).

\item Small-scale thread oscillations are detected in the quiescent
prominence, which were perpendicular to the prominence threads. They
are only pronounced in the H$\alpha$ line-center images and can last
for one entire cycle.

\item The thread oscillations exhibited two changing patterns. One showed
an initial period of $\sim$20~minutes, it firstly decreased with a
slowly rate and then it changed to increase. The other one exhibited
an initial period of $\sim$6~minutes, it grew quickly at the
beginning and then it changed to  decrease. It also exhibited
simultaneous drifting and oscillating motions.
\end{enumerate}

\acknowledgments We thank the anonymous referee for his/her valuable
comments and inspiring suggestions. The data used in this paper was
obtained by the NVST. The authors would like to acknowledge Dr.
L.~H., Deng and Y.~Y., Xiang for helping reconstructed data. This
work is supported by NSFC (Nos., 11603077, 11573072, 11773079,
11773068, 11790302, 11790300, 11729301, 11333009), the CRP
(KLSA201708), the Youth Fund of Jiangsu (Nos. BK20161095, and
BK20171108), the National Natural Science Foundation of China
(U1731241), the Strategic Priority Research Program on Space
Science, CAS (Nos., XDA15052200 and XDA15320301), and the Youth
Innovation Promotion Association of Chinese Academy of Sciences
(No., 2014047). D.~Li and Y.~Shen are supported by the Specialized
Research Fund for State Key Laboratories. The Laboratory No.
2010DP173032. Li \& Ning acknowledge support by ISSI-BJ to the team
of "Pulsations in solar flares: matching observations and models".

\begin{table*} \caption{The fitting values of the thread oscillations.} \centering
\begin{tabular}{c c c c c c c c }
 \hline\hline
Case &  A$_0$ (km)  & A$_m$ (km)  &  P$_0$ (min) & $\phi$    & k$_0$ (km~s$^{-1}$) & k$_1$ & k$_2$   \\
  \hline
1    &   1900       &    800      &    20    & 103$^{\circ}$ &   0   & -0.9 & 0.022          \\
2    &   1800       &    900      &    6     & -23$^{\circ}$ & 0.33  & 0.9 & -0.025          \\
\hline
\end{tabular}
\label{tab}
\end{table*}

\end{document}